\documentclass[iop]{emulateapj}
\newcommand{\beq}{\begin{equation}}
\newcommand{\eeq}{\end{equation}}

\newcommand{\citei}[1]{\citeauthor{#1} \citeyear{#1}}

\newcommand{\citeia}[2]{\citeauthor{#1}~(\citeyear{#1};~#2)}
\usepackage{color}
\usepackage{amsmath}
\slugcomment{}

\begin{document}
\title{Ultraviolet Extinction at High Galactic Latitudes}

\author{J.~E.~G.~Peek\altaffilmark{1}\altaffilmark{2}}
\author{David Schiminovich\altaffilmark{1}}
\altaffiltext{1}{Department of Astronomy, Columbia University, New York, NY. jegpeek@gmail.com}
\altaffiltext{2}{Hubble Fellow}

\begin{abstract}
In order to study the properties and effects of high Galactic latitude dust we present an analysis of 373,303 galaxies selected from the Galaxy Evolution Explorer (GALEX) All-Sky Survey and Wide-Field Infrared Explorer (WISE) All-Sky Data Release. By examining the variation in aggregate ultraviolet colors and number density of these galaxies we measure the  extinction curve at high latitude. We additionally consider a population of spectroscopically selected galaxies from the Sloan Digital Sky Survey (SDSS) to measure extinction in the optical. We find that dust at high latitude is neither quantitatively nor qualitatively consistent with standard reddening laws. Extinction in the $FUV$ and $NUV$ is $\sim$10\% and $\sim$35\% higher than expected, with significant variation across the sky. We find that no single $R_V$ parameter fits both the optical and ultraviolet extinction at high latitude, and that while both show detectable variation across the sky, these variations are not related. We propose that the overall trends we detect likely stem from an increase in very small silicate grains in the ISM.
\end{abstract}


\section{Introduction}\label{intro}

The interstellar medium (ISM) is suffused with astrophysical dust, a fine particulate material ranging   from plentiful large molecules up to rare micron-sized grains \citep{2003ARA&A..41..241D}. This dust is important both for the physics of the formation of planets and stars in the modern universe, as well as mediating the thermal characteristics of the ISM. It also plays a crucial role in our observations of everything beyond the solar system, affecting light through scattering, absorption, and emission from radio wavelengths through X-rays. Understanding the distribution, structure, and properties of dust is thus a central problem in modern astronomy. 

Extinction by dust at low opacity ($A_{\rm V} < 1$) is especially important for the study of objects beyond the Milky Way. We are now observing the cosmos with extremely high photometric precision in regions of the sky with low extinction. We have a particular interest in the effects of high-latitude dust in the context of large-area surveys, as they cover many disparate regions of the sky but aim to compile consistent samples of objects. Specifically, errors in dust extinction can have disastrous effects on the determination of cosmological parameters (e.~g.~ Huterer, Cunha, \& Fang, 2013, \citei{MKS10}). Secondarily, the study of diffuse dusty gas outside of galaxies is being explored for the first time, demanding a better understanding of how the observed extinction compares to the actual quantities of dust in low density environments (\citei{menard10}, \citei{2012arXiv1204.1978M}). More detailed studies of dust in these low-density regimes are crucial for these studies. 

A great advance was made by \citeia{SFD98}{SFD} who provided a Galactic dust map, using all-sky surveys in the far infrared by IRAS and DIRBE. Under the assumption of a single, consistent grain population, SFD employed the optically thin 100 micron flux from thermally emitting dust grains to estimate the column of dust along the line of sight, correcting for variation in observed temperature in the dust. Recently, there has a been a burst of activity mining Sloan Digital Sky Survey \citep{york00} to make data sets of ``standard crayons'' (objects whose observed aggregate color variations indicate foreground reddening) to study dust at high Galactic latitude and test the accuracy of SFD. \citet{Schlafly10} used the blue tip of the stellar locus, \citeia{2010ApJ...719..415P}{PG10} used spectroscopic observations of quiescent non-starforming galaxies, \citet{Jones:2011tf} used spectroscopic observations of M-dwarfs, and \citeia{2011ApJ...737..103S}{SF11} used spectroscopic observations of a broad range of stars. All of these explorations found that SFD is largely accurate in the optical, but that overall errors and spatial variability of the reddening are detectable. In particular, SF11 found that SFD overestimates reddening by 14\% and that the reddening law of \citeia{1999PASP..111...63F} {F99} is favored over \citeia{1994ApJ...422..158O}{O'D94} or \citeia{1989ApJ...345..245C}{CCM}.

We extend these explorations of high-latitude extinction into the ultraviolet using GALEX observations of galaxies as our standard crayons. The ultraviolet is an inherently more sensitive waveband for the study of dust, as it is typically $\sim3$ times more extinguished than the optical. Furthermore, the GALEX near-UV channel ($NUV$) covers the 2175 \AA~ bump, possibly a probe polycyclic aromatic hydrocarbons (PAHs; \citeauthor{Steglich:2010dy}~\citeyear{Steglich:2010dy}). The GALEX far-UV channel ($FUV$) probes the far-UV rise of the extinction curve, associated with very small silicate grains (\citeauthor{2001ApJ...548..296W}~\citeyear{2001ApJ...548..296W}; WD01). Small grains dominate the heating of the diffuse ISM \citep{Wolfire2003} and may act as tracers of variation in grain size distribution as a whole. Thus these grains are important both for the physics of the ISM, and better understanding the variation in grain size. CCM, O'D94, and F99 all report reddening laws in which the average UV extinction can be predicted from the overall dust column and a single parameter, $R_V \equiv A_V/E\left(B-V\right)$. These laws are derived by comparing stars extinguished by $A_V \gtrsim 1$ to relatively unextinguished stars, and examining the relative colors. It is standard practice to extrapolate these extinction laws to the vast areas of the sky of much lower extinction in which they have not been adequately tested. We rectify that problem here by measuring extinction at high latitudes directly.

In \S \ref{obs} we describe the observations and selection criteria that go into making our UV standard crayon sample of galaxies. In \S \ref{manda} we lay out how we grid the galaxy samples and how we compare these grids to Monte Carlo models. In \S \ref{results} we report the results of the analysis and provide sky maps of color, reddening and extinction. We discuss the implications of these results in \S \ref{discussion} and conclude in \S \ref{conclusions}. We include an Appendix that details some tests of systematic errors in our data and SFD.

\section{Observations and Data}\label{obs}

In addition to the published SFD reddening maps and the standard crayon catalog of quiescent galaxies from PG10, we use the full sky photometric catalogs from the GALEX satellite and the WISE satellite. 

\subsection{GALEX}
GALEX data were obtained as part of the GALEX All-Sky Survey (AIS) \citep{2005ApJ...619L...1M}, processed using the standard GALEX pipeline \citep{2007ApJS..173..682M} and released in catalog form as part of GCAT (Seibert et al.~, \emph{in prep}), based on the GALEX GR6 data release.  We selected all GALEX objects in the GCAT ASC table with $NUV <$ 20, a total of 7,577,668 objects with median exposure time in $FUV$ and $NUV$ of 144s.  The objects must be observed in GALEX $FUV$, although not necessarily detected. Because GALEX all-sky observations included in GR6 avoided the Galactic plane, these data only include a small number of objects ($<5\%$) at low Galactic latitude $|b|<10$. Standard GCAT aperture magnitudes and errors were used for $FUV$ and $NUV$.

Although we are using standard GALEX products we highlight one potential source of systematic error that could influence the results reported below.  We are most concerned with any systematic in $FUV-NUV$, which could result from a zero-point offset in either band.  Using WD standards, \citep{2007ApJS..173..682M} report random magnitude errors $\pm0.05, \pm0.03$ for $FUV$, $NUV$ respectively, and drifts $\sim0.015$ yr$^{-1}$ and systematic corrections between releases  $\sim0.01-0.02$.  This systematic is smaller than the effects we describe below by factors of a few, suggesting that future re-calibrations of the GALEX zero-point could have a quantitative but probably not a qualitative impact on the results we present here.

% DS 
% Make sure to include some discussion of calibration zero-point systematic between FUV and NUV.

\subsection{WISE}
We obtain Wide-Field Infrared Explorer \citep[WISE;][]{2010AJ....140.1868W} catalogs from the All-Sky Data Release (12 March 2012) Source Catalog archived at IRSA (NASA/IPAC Infrared Science Archive).  This catalog contains photometry for objects detected as a point source  (5$\sigma$) in any of the four WISE bands (WISE 3.4, 4.6, 12, 22 micron bands; W1, W2, W3 and W4).  We extract WISE photometric measurements for objects with positions within 4 arcseconds of the positions of GALEX $NUV$-detected objects described above.      We require the galaxies to be detected in W1, W2, W3 in areas of sky observed $\ge 8$ times ($>$ 60 s), the nominal WISE All-Sky Survey depth.    The WISE PSF is comparable to that of GALEX, $\sim$ 6 arcsecond (FWHM). All WISE photometry discussed in this paper are from WISE pipeline profile-fit fluxes.  We assume that all of our objects are unresolved by WISE, and have verified that our results are insensitive to this assumption.

% DS 
%  Currently WISE coverage map for W3 is intractable

\subsection{Galaxy Samples}\label{galsamp}
We use two galaxy samples to act as standard crayons, which allow us to probe high-latitude Galactic extinction along the line of sight in both the optical and the ultraviolet.

\begin{figure}
\includegraphics[scale=0.4]{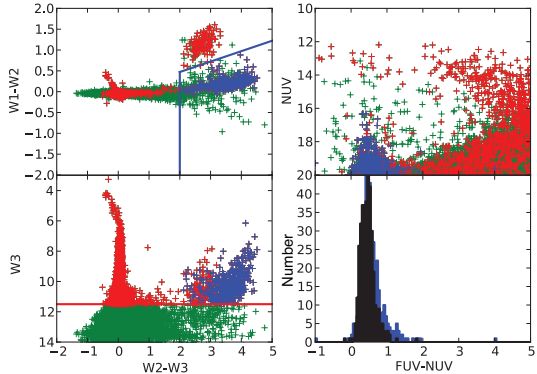}
\caption{Selection criteria for the GALEX-WISE galaxy sample. Green points represent the entire sample of objects matched between GALEX AIS and the WISE All-Sky survey. Red points represent objects with W$3 < 11.5$ (red line cut shown in lower left panel), blue points further meet the W3 - W2 $> 2$ and $\rm \left(W2- W3-3.15\right)/2 > \left(W1-W2-0.76\right)$ criteria (blue line cut shown in upper left panel). The lower right shows the similarity between the resultant galaxy sample (blue) and the SDSS MGS sample (black), in terms of $FUV-NUV$ color. Only small, randomly selected sub-samples are shown to reduce confusion, and match the sample sizes in the bottom-right panel.}
\label{cuts_wise_galex}
\end{figure}

The first galaxy sample we develop is taken from the WISE-selected sample of objects that appear in the GALEX AIS photometric data set (Figure \ref{cuts_wise_galex}).  We then make color and magnitude cuts based on the WISE and GALEX data to select for galaxies (see Figure \ref{cuts_wise_galex}). Firstly, we exclude objects with $NUV > 20$, as their color distribution is too broadened by noise to be useful at the depth of the AIS. A magnitude cut at W$3 < 11.5$ finds only objects with SNR $> 8$ in the W3 band. A color cut at W3 -- W2 $> 2$ removes some stars from the sample and a color-color cut of $\rm \left(W2- W3-3.15\right)/2 > \left(W1-W2-0.76\right)$ eliminates quasars \citep{Wu:2012ud}. The final GALEX-WISE galaxy sample includes 373,303 galaxies. We note that none of our IR magnitude or color cuts are dependent upon foreground extinction, as extinction in the mid-IR at the latitudes where AIS data are taken is negligible. As a check, we compare the ultraviolet color of these galaxies with GALEX AIS objects with $NUV < 20$ matched to the full spectroscopic SDSS MGS. These SDSS galaxies show an almost identical distribution in $FUV-NUV$ color. When visually examining SDSS photometry of the subset of GALEX-WISE objects that reside in the SDSS photometric area we find only slightly extended objects consistent with galaxies. It is simple to take the complementary WISE color-color cut ($\rm \left(W2- W3-3.15\right)/2 < \left(W1-W2-0.76\right)$) to generate a sample of 95,846 quasars. While this sample is smaller and somewhat broader in $FUV-NUV$ color, it serves as a sanity check against the GALEX-WISE galaxy sample.

The second data set is essentially identical to the 151,637 galaxies described in PG10. These galaxies are spectroscopically selected from the SDSS Main Galaxy Sample \citep{strauss02} Data Release 7 \citep{adelman-mccarthy08} to have neither detectable H $\alpha$ nor [O\textsc{ii}], and thus be non-starforming, using the values from the NYU Value Added Catalog \citep{blanton05-vagc}. Their colors are then corrected for the color-magnitude relation and color-density relations, as functions of redshift. As a result, their Galactic reddening-corrected colors have a extremely narrow distribution, with typical scatter of only 25 millimagnitudes in $g-r$ color. See PG10 for the details of this method. The only way the data set we use differs from the PG10 galaxies is that in this work we examine the colors of the galaxies uncorrected for Galactic extinction.

%FUV/SFD E(B-V) = 6.761, NUV/SFD E(B-V) = 6.857.

\section{Methods \& Analysis}\label{manda}

Our goal is to investigate the dependency of ultraviolet and optical reddening on the characteristics of the intervening gas, specifically the expected reddening along the line of sight determined from the far IR emission of the grains, as measured by SFD. This grain column is reported by SFD98 as an optical reddening, $E(B-V)$; SF11 demonstrated that the actual value of the reddening in these bands may be somewhat smaller than was originally proposed in SFD. To avoid confusion, throughout this work we will refer to the original value reported by SFD as $E_{\rm SFD}$, which readers might find easier to interpret as an extinction, $E_{\rm SFD} = A_{v, \rm SFD}/3.1$. To compare our observed extinction and reddening to extant reddening laws we use a modeled spectral energy distribution (SED) for a collection of 4000 galaxies that meet our selection criteria described in \S \ref{galsamp}, which has a very modest effect as compared to a simply flat SED \citep{2008MNRAS.388.1595D}.

In addition to simply determining the aggregate extinction and reddening parameters in the UV channels for the whole of the GALEX-WISE sky (\S \ref{vebv}) we wish to explore the variation and structure of ultraviolet extincting dust. We develop the necessary methods below.

\subsection{Gridding}\label{gridding}

Were our galaxies to be all identical copies, with zero intrinsic variation in color, and were our data to be noise-free, we could examine the relationship between color and $E_{\rm SFD}$ by simply showing a scatter plot. Nature is, of course, neither so simple nor so boring, and the intrinsic variation in both of our galaxy samples makes finding departures from the expected reddening along lines of sight to individual galaxies impossible. To explore the structure of ultraviolet extinction we  bin the data by position on the sky, making maps with pixels of 16 square degrees. This is a sensible approach in that it allows us to make maps of the color variation and provides a data set of points with well-determined $E_{\rm SFD}$, optical color, and UV color values. We use this pixel area to strike a balance between sensitivity and resolution.

To construct these maps we bin the data on a two Zenithal Equal Area grids, one for each Galactic hemisphere. An equal area projection is useful as it simplifies the significance variation in the data from pixel to pixel, and a projection center towards the zenith (and nadir, in the case of the Southern Galactic cap), reduces distortion in a map with little data in the Galactic plane. We note that it is more common to use Hammer-Aitoff projections for Galactic maps, but these tend to maximally distort at the poles. 

To determine the galaxy color assigned to that position in the map, we take the median of all galaxy colors within a pixel, $\tilde{C}_{a-b}$. We compute errors for each pixel in two ways and use the maximum of the two to be conservative. First, we compute a standard bootstrap error for the median color in each pixel, $\sigma_b$, resampling the galaxies with replacement. Second, we determine a typical width for the overall distribution by fitting Gaussian to the color histogram, $\sigma^\prime$, and generate errors for each pixel
\beq
\sigma_N = \sqrt{\frac{\pi}{2}}\frac{\sigma^\prime}{\sqrt{N}}.
\eeq
This method avoids problems with low-number statistics in bootstrap errors. There are two differences between the PG10 maps and the ones we make in this work. The pixels in our maps are statistically independent, and have no weighting function. The PG10 maps use a smoothing kernel to aid interpolation, and thus report a PSF-weighted median color and have covariant errors. This method can be applied to any quantity evaluated at the position of the galaxies. 
%At first examination, using this method on the $E_{\rm SFD}$ data would seem perverse to the point of diabolical: why take time-ordered data (IRAS \& DIRBE) and turn in into a map (SFD), only to evaluate said maps 
A key point is that we evaluate SFD maps at specific locations (the galaxy positions) and then rebuild maps from those locations. We choose to follow this procedure because it allows us to preserve the weighting in each pixel defined by the distribution of galaxies that contribute to the color map. If we were to use a map that was not based to the distribution of galaxies, we would be sampling observed reddening and $E_{\rm SFD}$ from slightly different areas of sky, which could lead to systematic errors, especially given the natural bias of a magnitude-limited sample of galaxies toward regions of lower extinction. We refer to this galaxy weighted median as $\tilde{E}_{\rm SFD}$. 

Implicit in our analysis is that galaxies act as identical objects; standard crayons in the terminology of PG10. While the selected galaxies vary significantly in color and many other properties, we are only concerned whether they intrinsically vary as a function of their position on the sky. In PG10 we found there was a slight trend that the selected galaxies were somewhat redder in denser environments, and corrected for this trend. Here we examine a similar metric, $\Sigma_5$, defined as the surface density of galaxies angularly closer than the 5th nearest neighboring galaxy. We evaluate this metric over low extinction regions of the sky, and find no detectable correlation between this measure and UV galaxy color. 

\subsection{Modeling}\label{modeling}

Maps of galaxy color and associated $E_{\rm SFD}$ allow us to make some inferences about the variation in optical and ultraviolet reddening across the high-latitude sky (see \S \ref{cmapssec}), but to learn about $R_{NUV} \equiv A_{NUV}/E_{\rm SFD}$ and $R_{FUV} \equiv A_{FUV}/E_{\rm SFD}$ directly, we must take into account both the color and the number density of galaxies on the sky. Note that here we are defining the $R_{UV}$s with respect to $E_{\rm SFD}$, rather than $E_{B - UV}$. To do this we construct a suite of simple but complete Monte Carlo simulations of the UV sky. For a choice of $R_{FUV}$ and $R_{NUV}$ we build a full mock distribution of galaxies on the sky, which we then grid as above in \S \ref{gridding}. First, we construct a subsample our 373,303 galaxies on the sky that have very low reddening, $0.02 < E_{\rm SFD} < 0.03$, which we refer to as the comparison sample (CS; see Appendix for discussion of errors in SFD at $E_{\rm SFD} < 0.02$). We use the CS, along with the intersection of the GALEX AIS exposure map and the WISE 8-exposure map, to determine the density of galaxies on the sky at low extinction, $n_g$. We then populate each map pixel area with galaxies drawn randomly from this distribution, matching the mean density observed at low extinction, as well as an empirically motivated scatter as described in Equation \ref{lss}, below. The large scale structure (LSS) of the universe causes significant variation in galaxy number counts than would be expected from simple Poisson noise. A full accounting of the 2-point (and N-point) correlation function of the galaxies in this sample is beyond the scope of this work, however, by examining the low-extinction region of the sky from which we drew the CS, we can determine the variation in galaxy counts as a function of the fraction of the sky observed in a given pixel. We find that the standard deviation in the number of galaxies in a pixel of area $\Omega$ is well fit by the simple function
\beq \label{lss}
\sigma_{g} = 0.37 n_{g} \left(\frac{\Omega}{\rm sq~ deg}\right)^{3/4}.
\eeq
While this method does not take into account any covariation in galaxy count of neighboring pixels, none of our results seem to depend on modeling this covariation.

Model galaxies are de-extinguished according to their original observed $E_{\rm SFD}$ and re-extinguished according to the $E_{\rm SFD}$ observed in their placed position and the chosen $R_{FUV}$ and $R_{NUV}$. The data are then gridded as described in \S \ref{gridding}. We also count the number of galaxies in each pixel $N_{NUV}$, and the number of galaxies in each pixel that have $FUV < 20$, $N_{\rm FUV}$, as a direct constraints on $R_{NUV}$ and $R_{FUV}$, respectively. We note that the exact way in which it is appropriate to de-extinguish the subsample galaxies is unclear. Is the $0.02 < E_{\rm SFD} < 0.03$ extinction reliable? Should we assume that $R_{\rm FUV}$ and $R_{NUV}$ are the canonical values or consistent across the sky? Is there a significant enough issue with the $E_{\rm SFD}$ zero-point such that de-extincting the galaxies at low extinction at all is too unreliable? While these effects do quantitatively effect our results weakly at very low extinction, they have no qualitative effect on our conclusions. 

The Monte Carlo model is then run 1000 times for each pair in a broad range of $R_{FUV}$ and $R_{NUV}$. We compare the observed values of $\tilde{C}_{FUV-NUV}$, $N_{NUV}$, and $N_{FUV}$ for each pixel to the cumulative distribution functions of these same statistics in a given Monte Carlo run. Taking into account covariance between these three statistics, we determine $\sqrt{ \chi_{red}^2}$ for each pixel for each model. This is equivalent to finding for each pixel where in the multivariate tolerance region of the model the observed $\tilde{C}_{FUV-NUV}$, $N_{NUV}$, and $N_{FUV}$ lie. Thus we determine what range of $R_{FUV}$ and $R_{NUV}$ provides an acceptable fit for each pixel in our maps.

%%%%% NOTE: how do we discuss non-Gaussianity? I think we are essentially assuming Gaussianity in our distribution, which is kind of a problem when we get to small #s. HALP.

\section{Results}\label{results}

\begin{figure}
\includegraphics[scale=1.0]{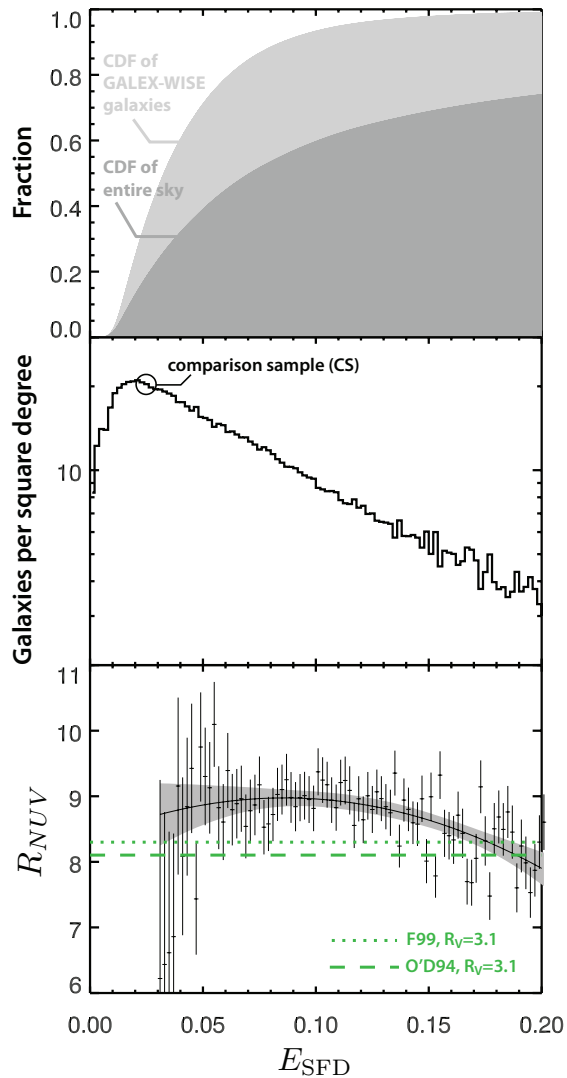}
\caption{The GALEX-WISE galaxy distribution as a function of $E_{\rm SFD}$. The top panel shows the cumulative distribution function (CDF) of the GALEX-WISE galaxy sample and the CDF of the whole sky, each as a function of $E_{\rm SFD}$. The middle panel shows the number density of the galaxies and the comparison sample, CS. The bottom panel shows the measured $R_{NUV}$ from the number density ratio to the CS and a second-order polynomial fit and 95\% confidence contours. Also shown is the expectation from O'D94 and F99 reddening laws.}
\label{vebv_ext}
\end{figure}

\begin{figure}
\includegraphics[scale=0.42]{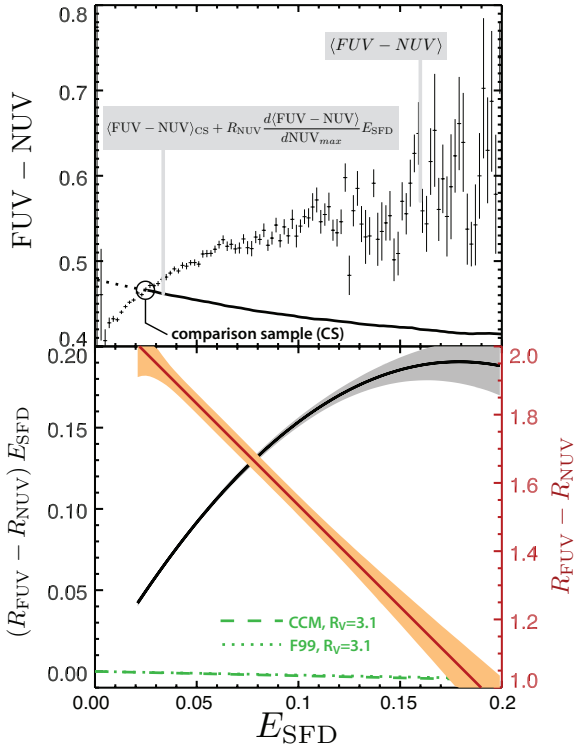}
\caption{Reddening measurement of the GALEX-WISE galaxies as a function of $E_{\rm SFD}$. The top panel shows the median color of galaxies as a function of $E_{\rm SFD}$, as well as the population reddening effect of Equation \ref{slope}. The bottom panel shows a second-order polynomial fit to the reddening corrected for the population reddening effect (see Equation \ref{fuvnuv}), representing the expected reddening of an object (black) and its derivative, the value of $R_{FUV} - R_{NUV}$ (red).}
\label{vebv_red}
\end{figure}

\begin{figure*}
\includegraphics[scale=0.97]{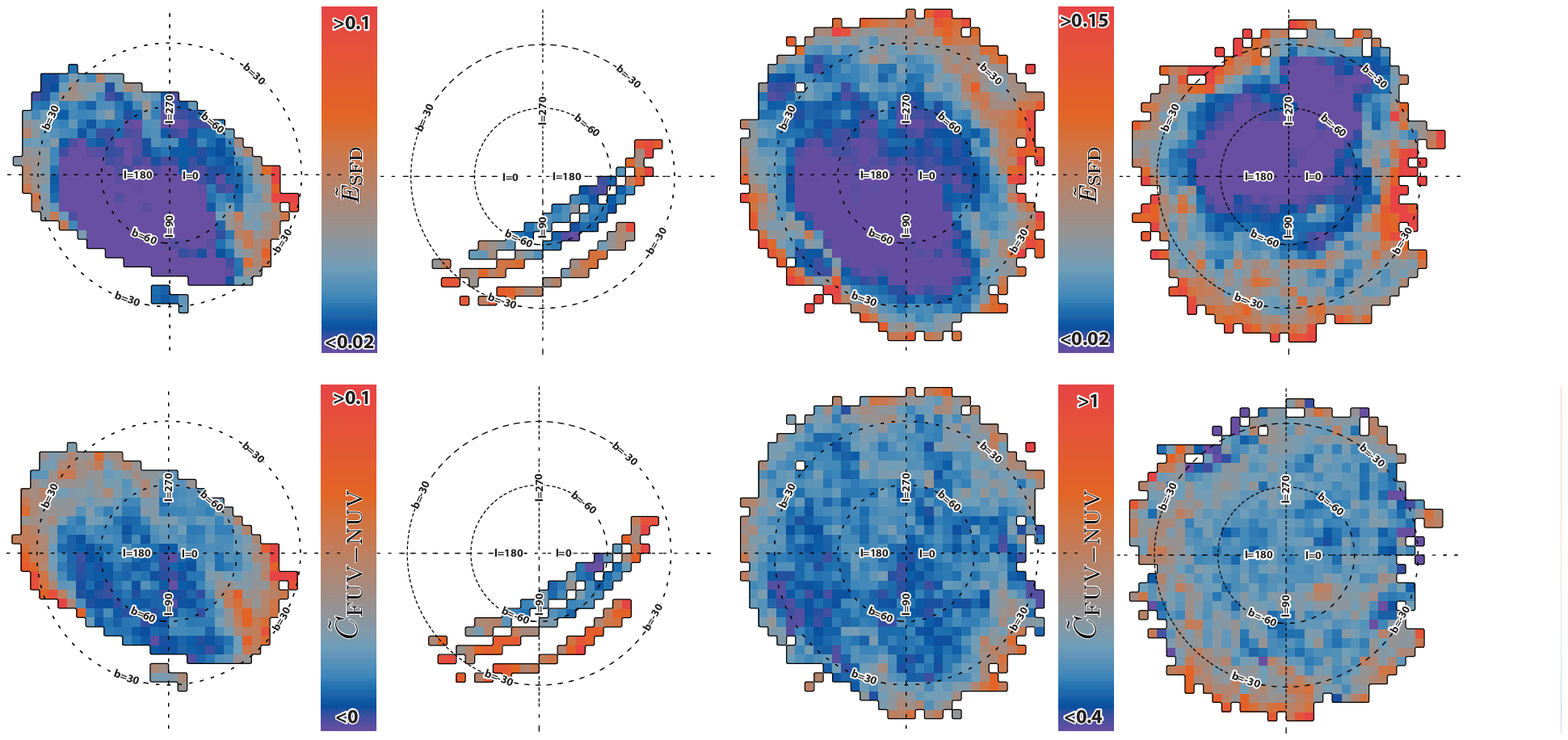}
\caption{Maps of $\tilde{E}_{\rm SFD}$ (top) and galaxy color (bottom). The right side columns show the PG10 galaxies, the left side columns show the GALEX-WISE galaxies. Maps are shown as a pair of hemispheres, Galactic North (left) and Galactic South (right). Maps are limited to the area in which the noise in the optical and UV color maps is less than 0.005 and 0.045 magnitudes, respectively.\vspace{15 pt}}
\label{cmaps}
\end{figure*}

\subsection{Overall Exctinction Parameters}\label{vebv}

To determine typical values for $R_{FUV}$ and $R_{NUV}$ over the whole of the GALEX-WISE sky, we simply bin galaxy data by $E_{\rm SFD}$, and measure colors and number counts of galaxies within those bins. We note that the area from which the GALEX-WISE galaxies were drawn represents nearly all of the sky at $E_{\rm SFD} \simeq 0$, dropping linearly to 50\% of the sky at $E_{\rm SFD} \simeq 0.2$. 

Two effects determine the dependence of typical color of background galaxies on $E_{\rm SFD}$ for a given reddening law. Primarily, the population of galaxies will redden (or `bluen') as $E_{\rm SFD}$ increases, dependent on the differences in extinction between the bands in question. There is a secondary effect that stems from the extinction of galaxies out of the sample population. If a population does not have a fixed color as a function of the magnitude on which it is selected, the typical color of the galaxies will change simply by removing the dimmest galaxies from the sample in regions of higher $E_{\rm SFD}$, which we call the population reddening effect. This is to say

\beq\label{slope}
\langle a-b\rangle = \langle a-b\rangle_0 + \left( (R_a-R_b) +R_c\frac{\partial\langle a-b\rangle}{\partial{c}_{max}}\right) E_{\rm SFD},
\eeq

\noindent where $\langle a-b\rangle$ is the observed color of a reddened galaxy population in $a-b$ color, $\langle a-b\rangle_0$ is the unreddened color, $R_a$ is the extinction in band $a$ per unit $E_{\rm SFD}$, and $c$ is the extinguished band upon which the sample is cut at $c_{max}$. Often, $R_a-R_b$ is of the order of $R_c$, and $\frac{\partial\langle a-b\rangle}{\partial{c}_{max}} << 1$. This is the case for the PG10 galaxies, which are measured in the optical and standardized to have fixed color with magnitude, and thus the population reddening effect is negligible. In the GALEX-WISE sample, $R_a-R_b = R_{FUV} - R_{NUV} \simeq -0.02$, $R_c = R_{NUV} \simeq 8$, and the population is uncorrected as a function of magnitude. Thus, the population reddening effect is non-negligible. In practice, we measure $\frac{\partial\langle a-b\rangle}{\partial{c}_{max}}$ by selecting a large number of galaxies from our sample with extremely low extinction, artificially extinguish them, re-apply the selection criterion, and measure the change in the median color of the population. We employ here the same comparison sample (CS) as described in \S \ref{modeling} ( $0.02 < E_{\rm SFD} < 0.03$).
%We call this population the comparison sample (CS) and define it to be all galaxies that meet the criteria $0.02< E_{\rm SFD} < 0.03$. We avoid including galaxies below $E_{\rm SFD} < 0.02$ due to issues with measured number densities in that regime (see Appendix). 

%Color errors are determined as a maximum of bootstrap and Poisson errors as in \S \ref{gridding}, number density errors taken to be Poisson.

Figure \ref{vebv_ext} shows the results of the extinction analysis. The $E_{\rm SFD}$ cumulative distribution function (CDF) in the top panel demonstrates that half of the entire sky has an extinction below $E_{\rm SFD} = 0.07$, showing the importance of extinction parameters at these low values of $E_{\rm SFD}$. We also show the CDF of the GALEX-WISE galaxies against $E_{\rm SFD}$, making plain the very small fraction of our 373,303 galaxies we observed towards $E_{\rm SFD} > 0.2$. The second panel shows the number density of galaxies in the sample, clearly showing the effects of extinction, along with the number density of the CS. The anomalous turn-over below $E_{\rm SFD} = 0.02$ is discussed in the Appendix. The third panel shows $R_{NUV}$ as a function of $E_{\rm SFD}$. These values are measured by comparing the number density of galaxies in the CS to the values within a given bin. A second-order polynomial fit is shown,
\begin{align}\label{rnuv}
&R_{NUV} = \left(8.36 \pm 0.42\right) + \nonumber \\
&\left(14.3 \pm 6.9 \right) E_{\rm SFD}+ \left(-82.8 \pm 26 \right) E_{\rm SFD}^2,
\end{align}
along with the 95\% formal confidence interval of the fit. 

Figure \ref{vebv_red} shows the results of the reddening analysis. The first panel shows the median color of the galaxies in each bin, the left hand side of Equation \ref{slope}. Errors are determined as a maximum of bootstrap and Poisson errors as in \S \ref{gridding}. Also shown is the population reddening effect we discussed above,  $\langle {FUV}-{NUV}\rangle_{\rm CS} + R_{NUV}\frac{\partial\langle {NUV}- {FUV}\rangle}{\partial{NUV}_{max}} E_{\rm SFD}$, as measured in the CS. We extrapolate this trend down to $E_{\rm SFD} = 0$. The second panel shows a second-order polynomial fit to 
\begin{align}\label{fuvnuve}
&\langle {FUV}-{NUV}\rangle - \langle {FUV}-{NUV}\rangle_{0} \nonumber \\
& - R_{NUV}\frac{\partial\langle {NUV}- {FUV}\rangle}{\partial{NUV}_{max}} E_{\rm SFD} = \nonumber \\
&\left(R_{FUV} - R_{NUV}\right) E_{\rm SFD} = \nonumber \\
&\left(2.11 \pm 0.08\right) E_{\rm SFD} + \left(-5.71 \pm 0.53 \right) E_{\rm SFD}^2,
\end{align}
which is equivalent to the expected color change for an individual object due to reddening.
We can determine the difference between the reddening coefficients, as shown in the bottom panel in red:
\begin{align}\label{fuvnuv}
&R_{FUV} - R_{NUV} =  \nonumber \\
&\left(2.11 \pm 0.08\right) + \left(-5.71 \pm 0.53 \right) E_{\rm SFD}.
\end{align}
Both fits are shown with 95\% confidence intervals. Also shown are the expectations from O'D94 and F99 extinction laws. Taking the sum of Equation \ref{rnuv} and Equation \ref{fuvnuv} we find
\begin{align}\label{rfuv}
&R_{FUV} = \left(10.47 \pm 0.43\right) + \nonumber \\
&\left(8.59 \pm 6.9 \right) E_{\rm SFD}+ \left(-82.8 \pm 26 \right) E_{\rm SFD}^2.
\end{align}

As a sanity check we apply the same analysis to the sample of 95,846 QSOs selected in \S \ref{galsamp}. These results are in agreement with the reddening analysis of the GALEX-WISE galaxy sample, reproducing both the shape and amplitude of the curves shown in the bottom panel of Figure \ref{vebv_red}, albeit with larger scatter.

Neither our observed extinction nor reddening are consistent with the literature. The standard values for extinction are $R_{NUV} = 8.3, 8.1$ and $R_{FUV} - R_{NUV} = -0.026, -0.022$ for the typical $R_V=3.1$ in the O'D94 and F99 extinction laws, respectively. Our values for $NUV$ extinction are inconsistent for the vast majority of our sample, and our $FUV-NUV$ reddening values are inconsistent throughout.  We note that neither of these values is consistent with the recalibration of SFD proposed by SF11, who predict a value of $R_{NUV} = 6.86$. SF11 note that for $E_{\rm SFD} < 0.2$, the SFD normalization is detectably higher, closer to the original value proposed in SFD, but do not find any evidence for a normalization greater than unity that would be needed to produce the offset we see in $R_{NUV}$. 

\subsection{Color Maps}\label{cmapssec}
Figure \ref{cmaps} shows the median color $\tilde{C}_{FUV-NUV}$, and $\tilde{E}_{\rm SFD}$ maps generated the method described in \S \ref{gridding} for the UV galaxies. $\tilde{C}_{FUV-NUV}$ varies dramatically over the map, from 0.35 to $> 0.8$. There does not seem to be any clear correlation between  $\tilde{C}_{FUV-NUV}$ and $\tilde{E}_{\rm SFD}$ except that redder UV colors only occur in regions of high extinction, whereas bluer colors can be found in regions of both high and low extinction. %In Figure \ref{evscolor} we show $\tilde{E}_{\rm SFD}$ vs. $\tilde{C}_{FUV-NUV}$ for all pixels in the UV maps. Along with confirming the overall picture shown in Figures \ref{vebv_ext} and \ref{vebv_red}, Figure \ref{evscolor} shows that there are a handful of highly reddened pixels, and very small number of pixels with negative reddening that are statistically significant.

\subsection{Contrast with Optical Reddening}\label{comparison}

%\begin{figure}
%\includegraphics[scale=1]{ebv_v_color}
%\caption{$\tilde{E}_{\rm SFD}$ vs. $\tilde{C}_{\rm FUV-NUV}$ for all pixels in the GALEX-WISE maps. The maximum of bootstrap errors and standard Poisson errors are shown for $\tilde{C}_{\rm FUV-NUV}$. The main features of interest are a widening cone of color toward increasing extinction, and a range of very red pixels.}
%\label{evscolor}
%\end{figure}

\begin{figure*}
\begin{center}
\includegraphics[scale=1.2]{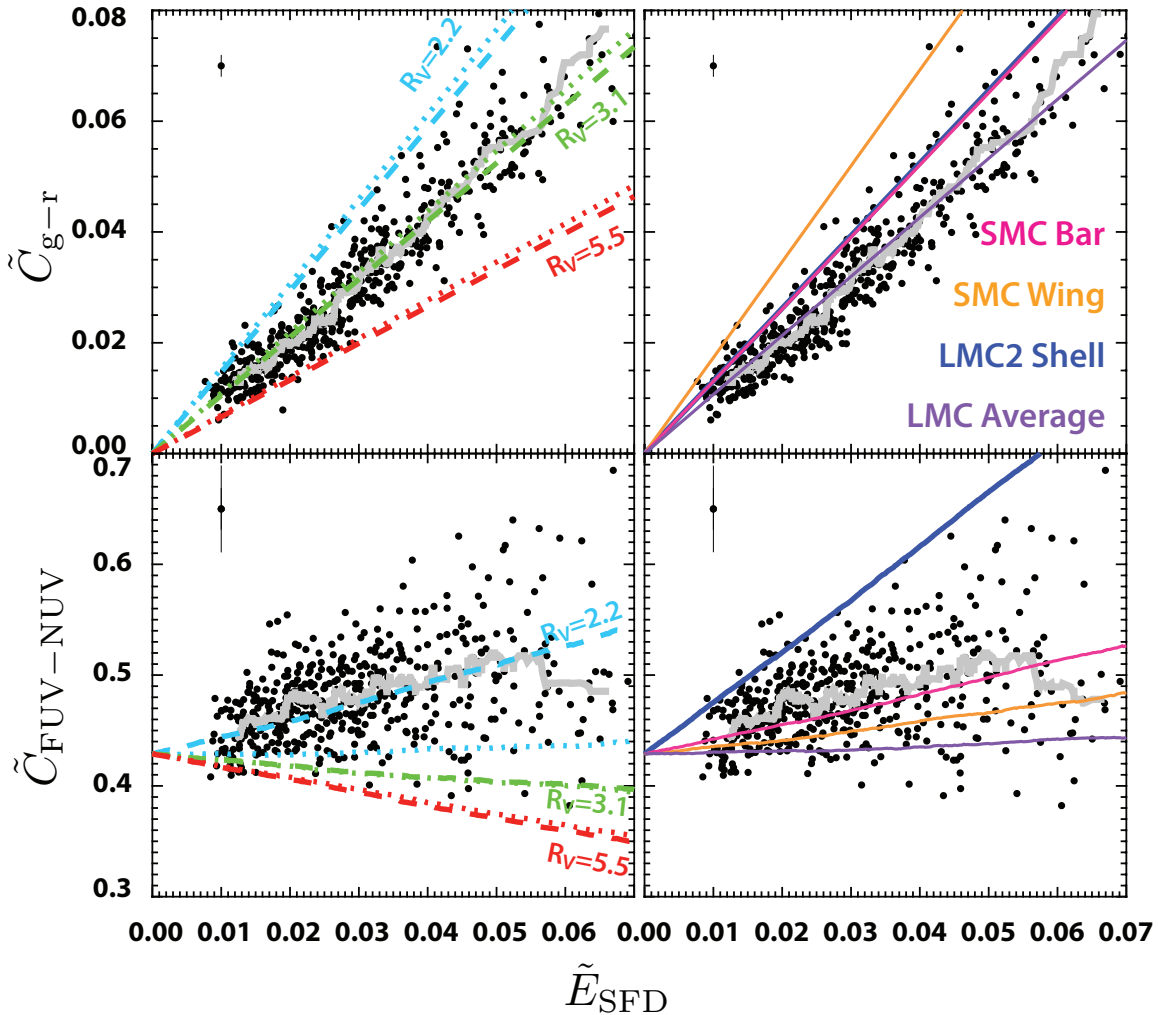}
\caption{$\tilde{E}_{\rm SFD}$ versus galaxy color, $\tilde{C}$ . Top left panel shows data from the GALEX-WISE galaxy sample, bottom left shows data from SDSS PG10 galaxy sample, both restricted to the SDSS footprint, for comparison. The gray lines show the median galaxy color as a function of $\tilde{E}_{\rm SFD}$. Also shown are O'D94 (dotted) and F99 (dashed) reddening curves with $R_V$ = 2.2, 3.1, and 5.5 (blue, green, and red, respectively). Neither reddening law is capable of reproducing the color variation in both the optical and UV. On the right the same data are shown with four Magellanic Cloud extinction curves from \citep{2003ApJ...594..279G}. Similarly, none of these extinction curves match the data. }
\label{sdss_a}
\end{center}
\end{figure*}

    Reddening laws make predictions for the extinction as a function of wavelength as a function of some measure of total dust column (here we use $E_{\rm SFD}$), and some other parameter that characterizes the grain distribution, typically $R_V$. To compare to reddening laws from the literature, we restrict our examination to the area of sky where we have both UV and optical extinction measures, the SDSS area. We show the distribution of  $\tilde{E}_{\rm SFD}$ against galaxy extinction for both UV and optical in Figure \ref{sdss_a}. In the left panel we show the expected reddening for these wavebands, using both O'D94 (nearly equivalent to CCM in these filters) and F99 reddening laws and a range of $R_V$. %We use a modeled spectral energy distribution for a collection of 4000 galaxies that meet our selection criteria described in \S \ref{galsamp}, which has a very modest effect on the expected reddening from these two laws. 
It is clearly evident that the canonical $R_V=3.1$, broadly consistent with the optical data using either reddening law, cannot explain the distribution of galaxy colors in the UV. The UV over this area is best fit with a F99 reddening law with $R_V \simeq 2.2$. As a further check we compare to reddening laws in the Magellanic Clouds as measured by \citep{2003ApJ...594..279G}. None of these reddening laws fit our data either, as shown in the right panels of Figure \ref{sdss_a}. Finally, we compare the high latitude UV extinction properties to each of the reddening laws toward 417 stars as measured by \citep{2004ApJ...616..912V}. Of these 417 stars, only two marginally match the measured $R_{FUV}$ and $R_{NUV}$ as measured at a typical high latitude value of $E_{\rm SFD} = 0.05$ (Equation \ref{rnuv} and Equation \ref{fuvnuv}). These stars, HD 13659 and HD 204827, have measured $R_V = 2.53, 2.58$, respectively, once again inconsistent with well-tested value of $R_V=3.1$. Thus, high latitude extinction is qualitatively inconsistent with typical reddening laws in the Galaxy, the Magellanic Clouds, as well as sightlines to many individual Galactic stars.

\begin{figure}
\includegraphics[scale=1]{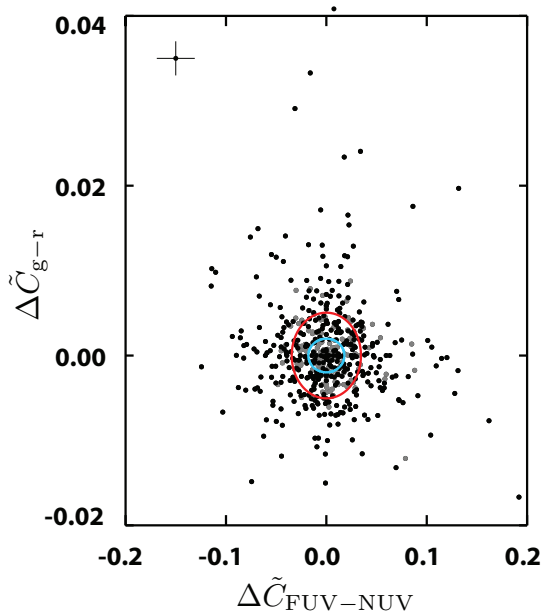}
\caption{The deviation of $\tilde{C}_{\rm g-r}$ vs. the deviation of $\tilde{C}_{FUV-NUV}$ over the SDSS DR7 footprint. Deviations are measured from the rolling median shown in gray in each panel of Figure \ref{sdss_a}. The red ellipse encloses 50\% of the data points, the blue ellipse would enclose 50\% of the data points were they all to be distributed from zero only according to their errors (i.~e.~ have a reduced $\chi^2 = 1$ to a model in which the data were simply scattered about the median according to their errors). The median error bar is shown in the upper left. We see that the scatter is significantly larger than would be expected from the error bars, and that there is no detectable correlation between $\Delta\tilde{C}_{\rm g-r}$ and $\Delta\tilde{C}_{FUV-NUV}$. This indicates that variation in reddening from the typical value in the UV is not primarily driven by the same ISM process that creates the variation in the optical reddening.}
\label{deltadelta}
\end{figure}

It was shown in PG10 that the deviation in $\tilde{C}_{g-r}$ from the expected value was significant in many regions of the SDSS footprint. To investigate the relationship between this variation and the equivalent variation in the UV, we show in Figure \ref{deltadelta} the residual $\tilde{C}$ for the GALEX-WISE and PG10 galaxies after subtracting the median trend with $\tilde{E}_{\rm SFD}$ shown in Figure \ref{sdss_a}. We find that the deviation in the UV color, $\Delta \tilde{C}_{FUV-NUV}$, is also in excess of what would be predicted from the errors, and that there is no detectable correlation between $\Delta \tilde{C}_{FUV-NUV}$ and $\Delta \tilde{C}_{\rm g-r}$. This indicates that whatever ISM process is driving variation in $R_{FUV} - R_{NUV}$ across the sky is not the same as the process driving $R_{\rm g} - R_{\rm r}$. Specifically, it is not simply errors in $E_{\rm SFD}$ generating both effects --- the reddening curve at high latitude is both different than predicted by O'D94 and F99, and its variations also do not follow a simple one-parameter model. 

\subsection{Extinction Modeling}\label{extmodeling}

\begin{figure*}
\includegraphics[scale=0.4]{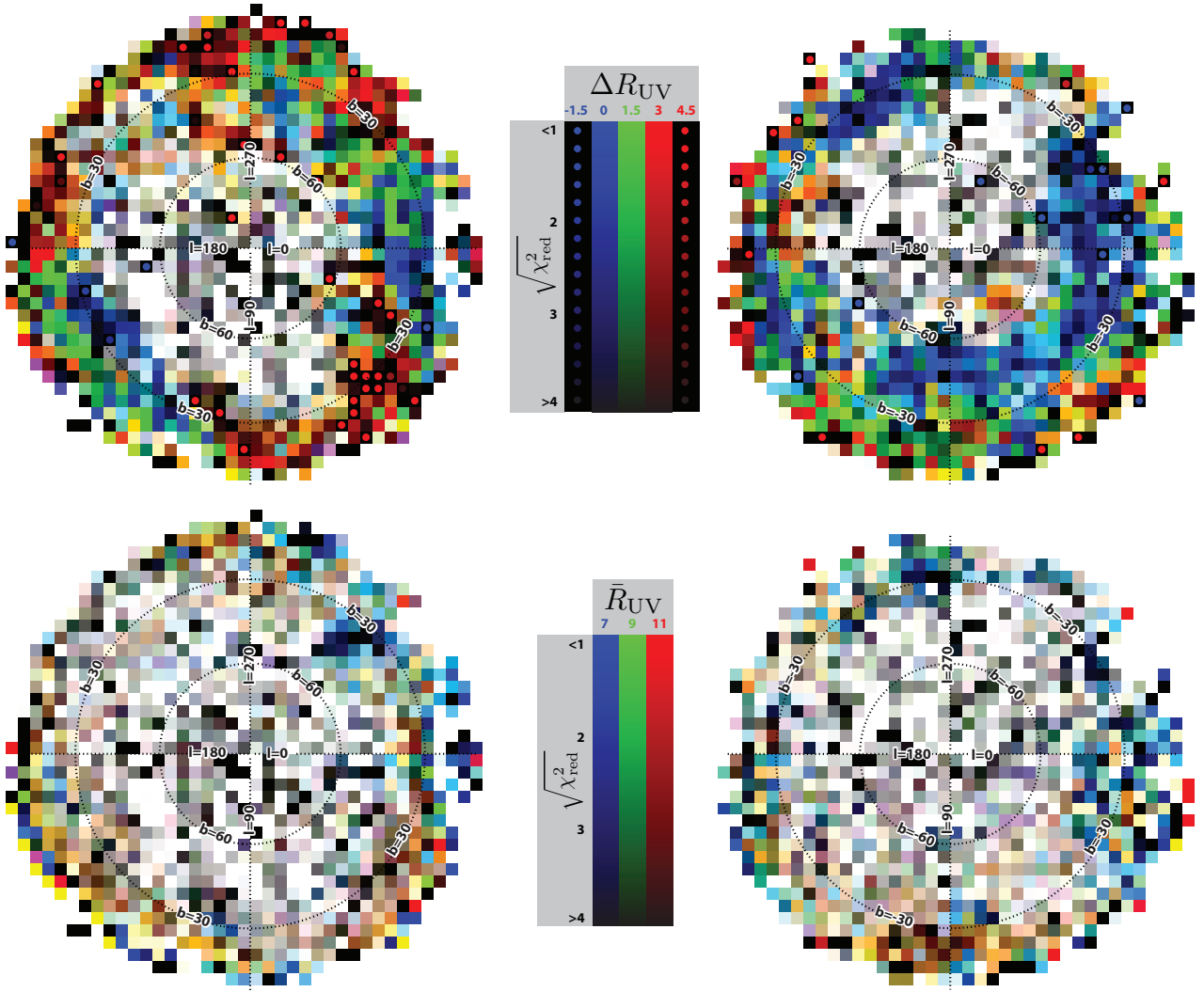}
\caption{%By examining the color and number density of 400,000 galaxies selected with WISE and GALEX, we are able to make estimates of the Galactic reddening parameters, $R_{\rm NUV} \equiv A_{\rm NUV}/E\left(B-V\right)$ and $R_{\rm FUV}$ using a montecarlo model. Above are 
Maps of the goodness-of-fit of each of our 15 Monte Carlo models. Top panels represent $\Delta R_{\rm UV} \equiv R_{FUV}-R_{NUV}$, bottom panels represent $\bar{R}_{\rm UV} \equiv \left(R_{FUV}+R_{NUV}\right)/2$;  Left panels show the northern galactic hemisphere, right panels show the southern Galactic hemisphere. The color of each 16 square degree region is colored according to the goodness-of-fit statistic, $\sqrt{\chi^2_{red}}$, it has for each model. In the top panels, red, green, and blue correspond to $\Delta R_{\rm UV} = \left[ 0, 1.5, 3 \right]$, with the brightness in each color representing the $\sqrt{\chi^2_{red}}$ in the best fitting model across the range of $\bar{R}_{\rm UV}$. In the bottom panels, red, green, and blue correspond to $\bar{R}_{\rm UV} = \left[ 7, 9, 11 \right]$, with the brightness in each color representing the $\sqrt{\chi^2_{red}}$ in the best fitting model across the range of $\Delta R_{\rm UV}$. Additionally, in the top panels, if the $\Delta R_{\rm UV} = \left[ 0, 1.5, 3 \right]$ models fail, each having a best-fit $\sqrt{\chi^2_{red}} > 4$, the best fit among $\Delta R_{\rm UV} = \left[ -1.5, 4.5 \right]$ is shown as a circle in blue or red respectively, with intensity of color once again representing goodness of fit.\\
The maps show that there are coherent radian-scale features in $\Delta R_{\rm UV}$ over large swaths of the sky, with a bias toward higher $\Delta R_{\rm UV}$ in the north and lower $\Delta R_{\rm UV}$ in the south. Models are less sensitive to variation in $\bar{R}_{\rm UV}$, but we do detect some large structures. Structures in the two metrics do not seem to be correlated.}
\label{MC_maps}
\end{figure*}

We have shown that reddening in the UV over the GALEX sky, $R_{FUV}-R_{NUV}$, does not follow the canonical laws in O'D94 or F99, and that variation in $R_{FUV}-R_{NUV}$ on the sky is neither simply a function of error in $E_{\rm SFD}$ nor driven by whatever grain population variation that generates the variation in the optical. To investigate the variation of UV extinction across the sky we model the full observed GALEX-WISE galaxy distribution using the method described in \S \ref{modeling}. We parameterize each model by $\Delta R_{\rm UV} = R_{FUV}-R_{NUV}$ and $\bar{R}_{\rm UV} = \left(R_{FUV}+R_{NUV}\right)/2$, as these quantities are largely independent; the first is largely driven by $\tilde{C}_{FUV-NUV}$ and the second by $N_{NUV}$ and $N_{FUV}$. After exploring the wide range of models we find we can fit the bulk of the data points with a grid of 15 models, five values of $\Delta R_{\rm UV} = \left[-1.5, 0, 1.5, 3, 4.5\right]$ and three values of $\bar{R}_{\rm UV} =\left[7, 9, 11\right]$. The results of the modeling are shown in Figure \ref{MC_maps}. We find that there are coherent $\Delta R_{\rm UV}$ structures on the sky that span radians in extent. The northern hemisphere seems to have more regions of large $\Delta R_{\rm UV}$, while the southern sky has more regions of low $\Delta R_{\rm UV}$, with a number of regions showing consistent ``bluening''. At high latitude, where ${E}_{\rm SFD} \lesssim 0.04$, we cannot discern between $\Delta R_{\rm UV}$ models. We have less success modeling $\bar{R}_{\rm UV}$, as we are very susceptible to variation in galaxy counts due to LSS. We do find large regions of the sky somewhat more prone to lower $\bar{R}_{\rm UV}$ and higher $\bar{R}_{\rm UV}$, and these regions do not seem to have a direct correspondence to the  variations in $\Delta R_{\rm UV}$.

\section{Discussion}\label{discussion}

\subsection{Implications for UV measurements}

Given that all UV extragalactic studies require a Galactic reddening correction, we might ask what a correction to the standard prescription will have on these studies.  While many analyses will be insensitive to the small corrections we have measured, dust attenuation measures using $FUV-NUV$ colors \citep[e.g.,][]{2000ApJ...533..682C}, and counts analyses (e.g. number counts, correlation functions) will often be quite sensitive to Galactic corrections on the scale we have detected.   

Dust attenuation measurements using the UV SED slope follow the spirit of the \citet{1999ApJ...521...64M} prescription, and in general find an approximate dependence $A_{FUV}\sim 4(FUV-NUV)$ with considerable scatter in the average slope and in the intrinsic distribution of the relationship across the galaxy population \citep{2005ApJ...619L..55S}---despite the many caveats, this measure has been applied to UV luminous galaxies across a wide range of redshifts.  Our results imply a correction to $-\Delta$(FUV-NUV) $\sim$0.05-0.1 (for a typical extragalactic line of sight with E(B-V) $\sim$ 0.03-0.07) implying a systematic dust attenuation correction of $-\Delta A_{FUV,cor}\sim 0.2-0.4$ (with galaxies intrinsically bluer than previously determined).  The effect on counts and correlation function analyses will depend on the slope of the extragalactic counts and fluctuations in the Galactic dust column/extinction correction.  The impact of Galactic reddening has been discussed previously \citep{2005ApJ...619L..11X, 2007ApJS..173..494M, 2009ApJ...698.1838H, 2010MNRAS.406..803B, 2010ApJS..190...43H}.  Here we merely emphasize that the implied systematic correction will provide an additional source of hitherto uncorrected variance for the counts analyses and component of the measured power spectrum that remains highly correlated with Galactic structure.

Galactic reddening corrections may also impact QSO colors, which may then be interpreted as a variation in intrinsic QSO reddening, or absorption of UV photons by intergalactic HI or dust.  As discussed above, we have tested our methodology using a smaller QSO sample and find similar results as with our main galaxy sample.   Observed-frame UV measurements have been used to determine the rest-frame EUV/UV spectrum of QSOs \citep{1997ApJ...475..469Z, Telfer:2002kn, 2012ApJ...752..162S}, requiring accurate Galactic reddening corrections.  Precise measurements may be able to detect the "Lyman valley", Lyman continuum absorption resulting from the integrated effect of many low density absorbers.  \citet{1990A&A...228..299M} predict Lyman valley absorption at the $\sim 5$\% level (at 700\AA) out to $z\sim 1$, an amplitude comparable to the Galactic reddening corrections proposed here.  In the most comprehensive study of the EUV spectrum of QSOs to date, \citet{2012ApJ...752..162S} are unable to detect this valley, though just at the limits of their statistics and using standard Galactic corrections.  Improved reddening corrections will better constrain this valley and the slopes of the ultraviolet composite spectrum in the EUV.

\subsection{Implications for Galactic Dust}

Reddening in the UV is thought to be governed by two groups of particles; molecule sized carbonaceous grains including PAHs and very small silicate grains (VSGs; \citei{Weingartner:2001eb}, \citei{Desert:1990ua}, \citei{Zubko:2004dk}). While the details of the models vary, PAHs and associated grains typically have a stronger extinction in the $NUV$ (near the 2175 \AA~ bump) and VSGs have more extinction toward the $FUV$. Some models reflect an additional large grain component responsible for reddening in the optical and relatively gray extinction in the UV, while others lump all silicate grains together with a grain size distribution. As standard extinction laws of the Milky Way parameterized only by $R_V$ fail in the UV for the low extinction areas surveyed in this work (Figure \ref{sdss_a}), we are interested in which components of the physical dust models could be altered to match our observations. Specifically, we wish to match the results that 1) at high latitude extinction in the NUV is $\sim10\%$ larger than expected, extinction in the FUV is $\sim30\%$ larger than expected, and reddening in the optical is unchanged.

Modifying the strength of any large grain component will do relatively little to match our observations. While we do see a 10\% increase in $R_{NUV}$ at $E_{\rm SFD} < 0.15$, we see no change in optical reddening. Since larger grains strongly affect reddening in the optical, variation in the larger grains is unlikely to be the dominant effect. The carbonaceous--PAH component typically provides more extinction in the $NUV$ than the $FUV$ GALEX bands, thus variation in this component provides correlation between increased overall extinction and decreased UV reddening. As this is the opposite of the effect we see, variation in the carbonaceous--PAH is unlikely to be the major culprit. Varying the VSG component may be the best option to fit our observations, providing a positive correlation between reddening and extinction in the UV. We note that since the extinction dramatically rises from the $NUV$ to the $FUV$ (10\% to 35\%) as compared to standard Milky Way reddening laws, and the optical reddening is unaffected, there must be a relative variation in the smallest silicate grains compared to larger silicate grains, rather than an overall renormalization of the silicates. We conclude that there does not exist a universal grain size distribution for the Milky Way, even allowing for parameterization with $R_V$. 

It is perhaps not surprising that the lower density environments probed by these sightlines would harbor a different population of grains. Dust grains are known to grow in the ISM, and this growth is strong function of gas density \citep{Draine:1990uf}. Furthermore, grain are shattered in the ISM in shocks, including grain-grain and grain-ion collisions \citep{Tielens94}. This is to say that if grains have been outside of the molecular environment long enough to be processed by shocks, we would indeed expect there to be an excess of small grains as observed. This is also in qualitative agreement with the work of \citep{2011ApJ...735....6P} in the LMC. We note that grain-size variation in the MW beyond the $R_V$ prescription has been noted before, in precision analyses of individual sightlines toward denser regions (\citei{Larson:2000ey}, \citep{Clayton:2003de}).

We also would expect there to be significant variation in the the fraction of PAHs as we look toward low density environments. Observations of the LMC \citep{Paradis:2009wu} and the SMC \citep{2010ApJ...715..701S} both show a correlation between the fraction of dust in PAHs and the density of the environment being probed. In our own Galaxy \citep{GS06} showed that there is a significant decrease in molecular fraction at column densities represented by $E_{\rm SFD} < 0.1$. While VSG variation seems to dominate PAH variation in aggregate, we do see evidence for a variation in Figure \ref{MC_maps} that cannot be explained by only varying the VSG fraction. Were all variation to be limited to the VSG fraction, there would be a strong positive correlation between UV color (top of Figure \ref{MC_maps}) and UV extinction (bottom of Figure \ref{MC_maps}). As this is not the case, there must be some added variation in dust grain properties at intermediate latitudes with an anti-correlation between reddening and extinction. This effect is consistent with variations of PAH fraction across the high latitude sky.

By connecting the angular structure in the $\Delta R_{\rm UV}$ maps (Figure \ref{MC_maps}) to the angular structure of other tracers of ISM structure, we could get a better physical understanding of the drivers of grain shattering and PAH fraction variation. A brief visual inspection of the locations of the four high-latitude radio loops \citep{Haslam:1971vh}, the H-alpha sky \citep{Finkbeiner:2003ce}, the soft X-ray background sky \citep{Freyberg:1998wy}, and the HI sky as a function of velocity \citep{Kalberla05} shows no convincing association with the $\Delta R_{\rm UV}$ maps. It is not entirely clear what to make of this lack of association, other than it may help to rule out some very simple methods of grain destruction that might predict a correlation. A more detailed comparison would be valuable, but is beyond the scope of this work.

%\subsection{Implications for LISM structure}
% - JEGP
% - Gould belt associations? 
% - Loops I, II, III, IV?
% - comparison to X-ray, H-alpha (WHAM?), HI, Erika, masking maps [Ghosh,... Gorski, 2012]

\section{Conclusions}\label{conclusions}

Simply put, \emph{ultraviolet extinction at high latitude is neither quantitatively nor qualitatively consistent with the standard Galactic reddening laws}. More specifically:
\begin{enumerate}
\item{The overall ultraviolet extinction parameter $R_{NUV} \simeq 9$ for  $E_{\rm SFD} \simeq 0.1$, trending toward $R_{NUV} \simeq 8$ near $E_{\rm SFD} \simeq 0.2$ (see Equation \ref{rnuv}). The UV reddening can be parameterized as in Equation \ref{fuvnuv}, starting at $R_{FUV} - R_{NUV} \sim 2$ and dropping toward 0 as $E_{\rm SFD}$ increases. At low extinction this represents $R_{FUV}$ and $R_{NUV}$ at 1.3 and 1.1 times their fiducial values and 1.6 and 1.3 times the values suggested by the rescaling of SFD in SF11.}
\item{The premise that dust extinction across wavelengths can be parameterized along the line sight by a single parameter, $R_V$, is not consistent with observations at high latitude. We find that over the SDSS spectroscopic area optical data are consistent with $R_V = 3.1$. In the same region, the UV reddening is inconsistent with CCM or O'D94 reddening, and only consisting with F99 reddening for $R_V \simeq 2.2$. Furthermore, both optical and UV reddening depart from these $R_V$ values significantly over the observed area, but these variations are not correlated with one another.}
\item{The UV high latitude sky is neither consistent with a single reddening parameter, nor a range of reddening parameters characterized by $E_{\rm SFD}$. Instead, $\Delta R_{\rm UV} \equiv R_{FUV} - R_{NUV}$ ranges from -1.5 to 4.5, in radian-scale coherent structures across the sky. Variation in $\bar{R}_{\rm UV}$ is less clear, but there seem to be large, coherent structures that do not follow the structures in $\Delta R_{\rm UV}$.}
\item{Extragalactic studies that use $FUV-NUV$ colors as dust attenuation measures, counts and correlation analyses, QSO EUV spectral energy distribution and low redshift ($z<1$) IGM absorption measurements may be sufficiently sensitive to require consideration of the variation in Galactic extinction and reddening.}
\item{These results are consistent with low density, high latitude ISM having a typically higher fraction of very small silicate grains, which generate both higher extinction and stronger reddening in the GALEX UV channels.}

\end{enumerate}

\acknowledgements

This publication makes use of data products from the Wide-field Infrared Survey Explorer, which is a joint project of the University of California, Los Angeles, and the Jet Propulsion Laboratory/California Institute of Technology, funded by the National Aeronautics and Space Administration. The authors thank Karin Sandstrom, Eddie Schlafly, and Robert Goldston for helpful comments and Lauren Corlies for providing the UV galaxy SEDs. {\it Galaxy Evolution Explorer} is a NASA Small Explorer, launched in April 2003. We gratefully acknowledge NASA's support for construction, operation, and science analysis for the GALEX mission, developed in cooperation with the Centre National d'Etudes Spatiales of France and the Korean Ministry of Science and Technology. JEGP was supported by HST-HF-51295.01A, provided by NASA through a Hubble Fellowship grant from STScI, which is operated by AURA under NASA contract NAS5-26555. 

\begin{center} 
{\sc Appendix}
\end{center}

Our analysis presupposes that the positions of galaxies on the sky are not correlated with the foreground extinction in the Galaxy. While this is likely true in reality, throughout this work we use the SFD extinction map to determine foreground extinction, which may itself be incorrect. \citeauthor{Yahata07}~(\citeyear{Yahata07};Y07) purport to find just such a correlation which we must treat with great care. They showed that the number counts of galaxies found photometrically in SDSS DR4 were significantly lower than expected at very low $E_{\rm SFD}$. They develop a conjecture that the contribution of far-IR flux from the galaxies themselves create this effect by artificially increasing $E_{\rm SFD}$ along lines of sight to galaxies, thus decreasing the number of galaxies towards low extinction. The fact that galaxies contribute significantly to the SFD map is confirmed in \citet{Kashiwagi:2012we}.

\begin{figure}
\includegraphics[scale=1]{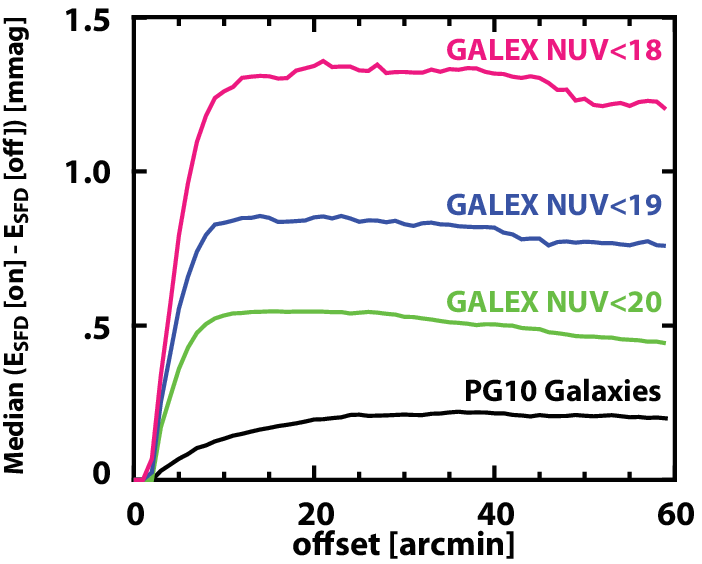}
\caption{The change in median value of $E_{\rm SFD}$ as we displace Galactic latitude away from our galaxies, $E_{\rm SFD}\left[l\right] - \left(E_{\rm SFD}\left[l+\delta l\right] + E_{\rm SFD}\left[l - \delta l\right]\right)/2$. We do detect contamination at the sub-millimagnitudesub-millimag level in both PG10 and GALEX-WISE galaxies. This effect is consistent with far-IR associated with the galaxies themselves, though we find it to be too small to account for the observed decrement of low $E_{\rm SFD}$ galaxies as proposed in \citet{Yahata07}.}
\label{offset}
\end{figure}

\begin{figure}
\includegraphics[scale=1]{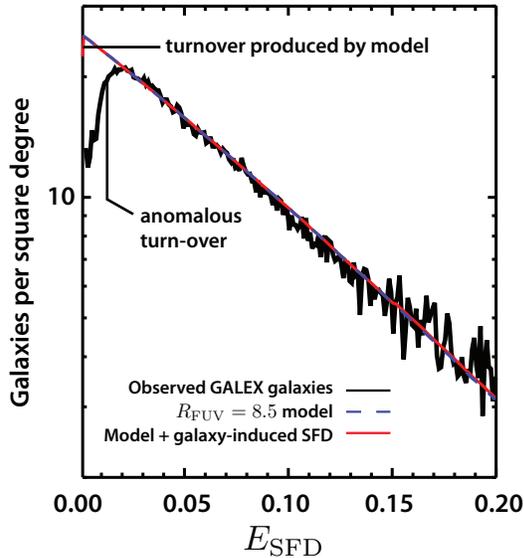}
\caption{Histogram of GALEX-WISE galaxy surface density as a function of $E_{\rm SFD}$. The black line is the observed distribution, showing an anomalous turnover at $E_{\rm SFD}=0.02$, as seen for galaxies in Y07. The dashed blue line is a simple model, extincting a selection of the GALEX-WISE population with $R_{NUV} = 8.3$. The red line shows the same model, but including additional $E_{\rm SFD}$ consistent with the measurements shown in Figure \ref{offset}. }
\label{y07model}
\end{figure}

To test this effect in our data, we compare the $E_{\rm SFD}$ toward our galaxies to the $E_{\rm SFD}$ for positions offset by $\pm\Delta l$. If the galaxies do indeed contribute significantly to $E_{\rm SFD}$, there should be a detectable increase in $E_{\rm SFD}$ towards the galaxies as compared to all the positions slightly offset in Galactic longitude. In Figure \ref{offset} we see just such an effect in both the PG10 and GALEX-WISE populations. The difference rises steeply within the $\sim 6^\prime~E_{\rm SFD}$ pixel, and flattens by about $\Delta l = 25^\prime$. The effect is weaker for the PG10 galaxies than the GALEX-WISE galaxies, and increases in strength when only a subset of the brighter GALEX-WISE galaxies is selected. All of these assessments are qualitatively in agreement with the conjecture in Y07 and quantitatively in agreement with the results of \citet{Kashiwagi:2012we}. 
o
To determine whether this effect is strong enough to produce the distribution of galaxies we observe, we build a simple model. We take the CS galaxies and artificially extinguish them, remove galaxies with $NUV >$ 20, and bin them by extinction. As shown in Figure \ref{y07model}, this model roughly reproduces the exponential slope in galaxy number density vs. $E_{\rm SFD}$ for $R_{NUV} = 8.5$. The model does not reproduce the anomalous turn-over at $E_{\rm SFD} = 0.02$ in our data, which is analogous to the effect observed first in Y07. We also show the same model, but now include the detected correlation between GALEX-WISE galaxies and excess $E_{\rm SFD}$ as a function of galaxy $NUV$ magnitude. While there is a marginal effect on the model at extremely low extinction, it is far too weak to reproduce the turnover. We conclude that the decrement of galaxies at low extinction, while clearly visible in our data, is not due to galaxy-induced variation in $E_{\rm SFD}$. We do not model any overall offset in the $E_{\rm SFD}$ zero-point that might stem from the aggregate of all galaxies. While such an effect could reduce the significance of the SFD signal toward low $E_{\rm SFD}$, and thus flatten the slope of galaxy number density histogram, we do not find any way that this effect can create a turn-over. It remains unclear what causes the decrement in galaxies where $E_{\rm SFD} < 0.02$. To avoid contaminating our analysis with this clear error, we ignore the few areas of sky where $E_{\rm SFD} < 0.02$ when galaxy number density is under consideration.

\

\begin{figure}
\includegraphics[scale=1]{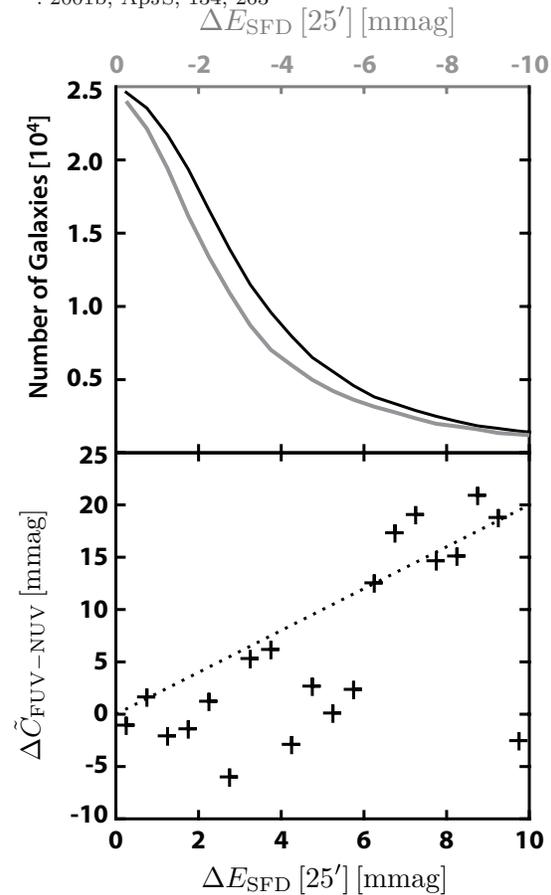}
\caption{The effect of $\Delta E_{\rm SFD}\left[25^\prime\right]$ on galaxy UV color. The top panel shows the distribution of GALEX-WISE galaxies IR excess as compared to positions offset by $25^\prime$. Black shows the distribution of excesses, gray the distribution of decrements. As expected, more galaxies have excess far-IR than decrement. The bottom panel shows the difference in color between galaxies with excess and decrement IR. We see a weak trend, with a slope no greater than about two. Overall, we estimate that galaxies will look artificially too red because of this effect by less than 0.6 millimagnitudes.}
\label{twop}
\end{figure}

Y07 is concerned with systematically incorrect foreground extinctions toward galaxies, and the impact of that error on the number density of background galaxies. Far more nefarious for our data is a known intrinsic dependence of galaxy UV color on far IR luminosity \citep{2000ApJ...533..682C}. Galaxies that are redder in the UV tend to have stronger far IR flux, which could bias our results if that same far IR light systematically contaminates $E_{\rm SFD}$. To address this effect we investigate the variation of median UV galaxy color as a function of the this erroneous IR component associated with the galaxies. We examine the distribution of $\Delta E_{\rm SFD}\left[25^\prime\right]$, the typical increase in $E_{\rm SFD}$ at a galaxy as compared to positions offset by $\pm 25^\prime$ in Galactic longitude. The distribution is shown in Figure \ref{vebv_red}. We then examine median galaxy color, $\tilde{C}$, as a function of $\Delta E_{\rm SFD}\left[25^\prime\right]$. We have found that galaxies typically become somewhat redder in the UV as $E_{\rm SFD}$ increases (see Figure \ref{vebv_red}), and since variability and amplitude in $E_{\rm SFD}$ are correlated, there is a spurious correlation between galaxy color and $\left|\Delta E_{\rm SFD}\left[25^\prime\right]\right|$. To nullify this effect we examine the difference in median color of galaxies between excess and decrement IR, 
\begin{align}
%\beq
%\begin{align}
\Delta\tilde{C}_{FUV -NUV}\left(\Delta E_{\rm SFD}\left[25^\prime\right]\right) \equiv \\
\tilde{C}_{FUV -NUV}\left(\Delta E_{\rm SFD}\left[25^\prime\right]\right) - \nonumber \\
\tilde{C}_{FUV -NUV}\left(-\Delta E_{\rm SFD}\left[25^\prime\right]\right). \nonumber
%\end{align}
%\eeq
\end{align}
In the bottom panel of figure \ref{twop} we show that there is a trend in galaxy UV color with $\Delta E_{\rm SFD}\left[25^\prime\right]$ of $\sim 2$ magnitudes per magnitude variation in $\Delta E_{\rm SFD}$. We find that a distribution of galaxies with this observed $\Delta E_{\rm SFD}$ and UV color trend only generates a typical shift in UV color of about 0.6 millimagnitudes, far below our sensitivity. 

%\end{appendix}
%\bibliographystyle{apj}

\end{document}